\documentclass[aps,prl,twocolumn,groupedaddress]{revtex4}

\usepackage{graphicx}% Include figure files
\usepackage{dcolumn}% Align table columns on decimal point
\usepackage{bm}% bold math

\begin{document}

\preprint{ELSAG Wigner rispedito.tex}

\title{{\bf Experimental Eavesdropping Attack against Ekert's Protocol based on Wigner's Inequality}}

\author{F. A. Bovino }%
 \email{Fabio.Bovino@elsag.it}
\author{A. M. Colla}
\author{G. Castagnoli}

\affiliation{%
Elsag spa \\
Via Puccini 2-16154 Genova (Italy)
}%

\author{S. Castelletto}
\author{I. P. Degiovanni}%
 \email{degio@ien.it}
\author{M. L. Rastello}

\affiliation{%
Istituto Elettrotecnico Nazionale G. Ferraris \\
Strada delle Cacce 91-10135 Torino (Italy)
}%

\date{\today}

\begin{abstract}

We experimentally implemented an eavesdropping attack against the
Ekert protocol for quantum key distribution based on the Wigner
inequality. We demonstrate a serious lack of security of this
protocol  when the eavesdropper gains total control of the source.
In addition we tested a modified Wigner inequality which should
guarantee a secure quantum key distribution.

\end{abstract}


\maketitle

Quantum key distribution (QKD) provides a method for distributing
a secret key for unconditional secret communications based on the
"one time pad" because it guarantees that the presence of any
eavesdropper compromising the security of the key is revealed. For
a review on this topic see \cite{gisinrevmod}.

The first protocol for QKD has been proposed in 1984 by Bennett
and Brassard \cite{bennet&brassard}, the worldwide famous BB84
protocol. In 1991 A. Ekert proposed a new QKD protocol whose
security relies on the non-local behavior of quantum mechanics,
i.e., on Bell's inequalities \cite{ekert}.

Several groups around the world implemented and tested QKD systems
based on variants of the BB84 protocol using either faint laser
\cite{faint1,faint2,faint3,faint4,faint5,faint6} or entangled
photons \cite{ekertrarity,sasha,qk2,qk3,qk1}, while, to our
knowledge, only recently two groups implemented the Ekert's
protocol \cite{qk2,qk3}. In particular Naik \textit{et al.}
\cite{qk3} implemented a variant of the Ekert's protocol based on
Clauser-Horne-Shimony-Holt (CHSH) inequality as proposed in
Ekert's paper \cite{ekert}, and Jennewein \textit{et al.}
\cite{qk2} implemented the Ekert's protocol based on the Wigner
inequality.

In ref. \cite{qk2} the Wigner inequality was first proposed to
provide an easier and equally reliable eavesdropping test as the
CHSH when the Ekert protocol is implemented. The necessary
security proof of the Ekert protocol based on the Wigner
inequality consists in verifying the violation of $W\geq 0$.

To obtain the Wigner inequality ($W\geq 0$) it is necessary to
review the Wigner argument \cite{wigner}. Two main assumptions are
stipulated in the proofs of the Wigner inequality: locality and
realism. Locality means that Alice's measurements do not influence
Bob's measurements, and \textit{vice versa}. Realism means that,
given any physical property, its value exists independently of its
observation or measurement. The counterpart of the local-realistic
theories is the non-locality behavior of quantum mechanics, a
signature of quantum entanglement. In particular Wigner considered
a quantum system prepared in the singlet state, and he obtained
the violation of the inequality $W\geq 0$, i.e. , $W=-0.125$.
Furthermore, in the derivation of his inequality, Wigner assumed
perfect anticorrelation in the measurement results. This
assumption is obviously reasonable in the test of realism and
locality of a physical theory (it reflects the classical
counterpart of a quantum system prepared in the singlet state).
Nevertheless, in terms of QKD this assumption corresponds to a
lack of security.

In fact, when the eavesdropper, Eve, measures photons on either
one or both of Alice and Bob channels, her presence should be
revealed by a higher value of $W$ than the local-realistic
theories limit $W=0$, as it happens for the CHSH inequality
\cite{ekert}. Unfortunately this is not the case. In fact, only
when Eve adopts an \textit{intercept-resend} strategy and detects
one photon of the pair, the inequality becomes $W \geq~$0.0625,
but, as we will show, this is not for eavesdropping on both
channels, because in this case there is no limit \cite{wignerien}.

In this letter we perform an experiment proving the weakness of
the Wigner inequality as a security test for QKD, under the
condition of Eve gaining total control of the source of photon
pairs. Under this condition, she prepares each particle of the
pair separately in a well defined polarization direction, in other
words she prepares the photon in Alice's channel in the state
$|\phi_{A} \rangle$, and the photon in Bob's channel in the state
$|\phi_{B} \rangle$, respectively
\begin{eqnarray*}
\left| \phi_{A}\right\rangle  &=&\cos \phi_{A}\left|
H_{A}\right\rangle +\sin \phi_{A}\left| V_{A}\right\rangle , \\
\left| \phi_{B}\right\rangle  &=&\cos \phi_{B}\left|
H_{B}\right\rangle +\sin \phi_{B}\left| V_{B}\right\rangle .
\label{evestat}
\end{eqnarray*}
Thus Eve has a perfect knowledge of the polarization of the
photons sent and, even if the non-local behavior of the original
quantum system (the singlet state) is completely removed, we prove
that she can avoid disclosing herself.

\begin{figure}[tbp]
%[htbp]
\par
\begin{center}
\includegraphics[angle=0, width=9 cm, height=6 cm]{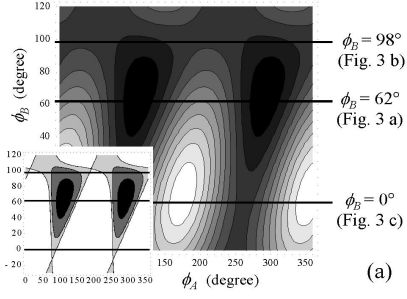}
\includegraphics[angle=0, width=9 cm, height=6 cm]{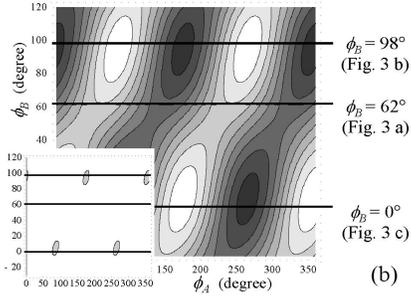}
\end{center}
\caption{ Contour-plot of $W$ (a) and $\widetilde{W}$ (b) versus
$\phi_{A}$ and $\phi_{B}$. Inset shows the regions where
$W<-0.125$ (black), $-0.125<W<0$ (dark grey), $0<W (\widetilde{W})
<0.0625$ (light grey), $W (\widetilde{W})>0.0625$ (white). }
\label{Figure 1}
\end{figure}

We remind the reader that ref. \cite{wignerien} presented a
modified version of the Wigner's parameter $\widetilde{W}$ which
maintains the same limits, i.e. , $\widetilde{W}\geq 0$ for local
realistic theories and $\widetilde{W}=-0.125$ for the singlet
state, but allows secure QKD, because $\widetilde{W}$ contains the
additional term accounting for the anticorrelation. In our
experiment we also measure $\widetilde{W}$ and we observe that the
minimum of $\widetilde{W}$ is well above the limit for
local-realistic theories in agreement with the theory
\cite{wignerien}, ensuring a secure QKD.

The measured quantities in our experiment are $W$ and
$\widetilde{W}$ \cite{wignerien}, respectively
\begin{eqnarray}
W &=&p_{-30^{\circ} _{A},0^{\circ} _{B}}(+_{A},+_{B})+p_{0^{\circ}
_{A},30^{\circ} _{B}}(+_{A},+_{B})   \label{WIG1} \\
&&-p_{-30^{\circ} _{A},30^{\circ} _{B}}(+_{A},+_{B}).  \nonumber
\\
\widetilde{W} &=&p_{-30^{\circ} _{A},0^{\circ}
_{B}}(+_{A},+_{B})+p_{0^{\circ}
_{A},30^{\circ} _{B}}(+_{A},+_{B})+   \label{WIG2} \\
&&p_{0^{\circ} _{A},0^{\circ} _{B}}(-_{A},-_{B})-p_{-30^{\circ}
_{A},30^{\circ} _{B}}(+_{A},+_{B}).  \nonumber
\end{eqnarray}
where $p_{\alpha _{A},\alpha _{B}}(x_{A},y_{B})$ are the
probabilities of detecting the pair of photons by the couple of
detectors $x_{A},\,y_{B}$ ($x_{A}=+_{A},-_{A}$ and
$y_{B}=+_{B},-_{B}$) when in the detection apparatuses two
half-wave plates (HWPs) project photons in the polarization bases
\begin{eqnarray*}
\left| s_{\alpha _{z}}\right\rangle  &=&\cos \alpha _{z}\left|
H_{z}\right\rangle +\sin \alpha _{z}\left| V_{z}\right\rangle , \\
\left| s_{\alpha _{z}}^{\bot }\right\rangle  &=&\sin \alpha
_{z}\left| H_{z}\right\rangle -\cos \alpha _{z}\left|
V_{z}\right\rangle , \label{polbas}
\end{eqnarray*}
with $z=A,B$. In formula
\begin{eqnarray}
p_{\alpha _{A},\alpha _{B}}(+_{A},+_{B}) &=&\left| \left\langle
\phi _{A}|s_{\alpha }{}_{A}\right\rangle \left\langle \phi
_{B}|s_{\alpha
}{}_{B}\right\rangle \right| ^{2}  \nonumber \\
p_{\alpha _{A},\alpha _{B}}(+_{A},-_{B}) &=&\left| \left\langle
\phi _{A}|s_{\alpha }{}_{A}\right\rangle \left\langle \phi
_{B}|s_{\alpha }^{\bot
}{}_{B}\right\rangle \right| ^{2}  \nonumber        \\
p_{\alpha _{A},\alpha _{B}}(-_{A},+_{B}) &=&\left| \left\langle
\phi _{A}|s_{\alpha }^{\bot }{}_{A}\right\rangle \left\langle \phi
_{B}|s_{\alpha
}{}_{B}\right\rangle \right| ^{2}      \nonumber    \\
p_{\alpha _{A},\alpha _{B}}(-_{A},-_{B}) &=&\left| \left\langle
\phi _{A}|s_{\alpha }^{\bot }{}_{A}\right\rangle \left\langle \phi
_{B}|s_{\alpha }^{\bot }{}_{B}\right\rangle \right| ^{2}
\label{prob}.
\end{eqnarray}

In Fig. 1 (a) and (b) we present the calculated contour plots of
$W$ and $\widetilde{W}$ versus the polarization directions
$\phi_{A}$ and $\phi_{B}$ of the photons of the pair sent by Eve.
Highest values of $W$ and $\widetilde{W}$ ($max(W)\simeq
max(\widetilde{W})\simeq 0.9557$) corresponds to the center of
white regions of Fig.s 1. Darker regions correspond to lower range
of values for $W$ and $\widetilde{W}$.

The values $min(W)\simeq -0.2121$ are in the middle of black
regions of Fig. 1 (a) along "Fig. 3 (a)" line, while
$min(\widetilde{W})\simeq 0.0443$ are almost in the middle of dark
grey regions of Fig. 1 (b). The straight lines for
$\phi_{B}=0^{\circ},62^{\circ}, 98^{\circ} $ represent sections of
the plots where the theoretical predictions are compared with the
experimental results of Figs. 3 (a), (b) and (c).

\begin{figure}[tbp]
%[htbp]
\par
\begin{center}
\includegraphics[angle=0, width=9 cm, height=6 cm]{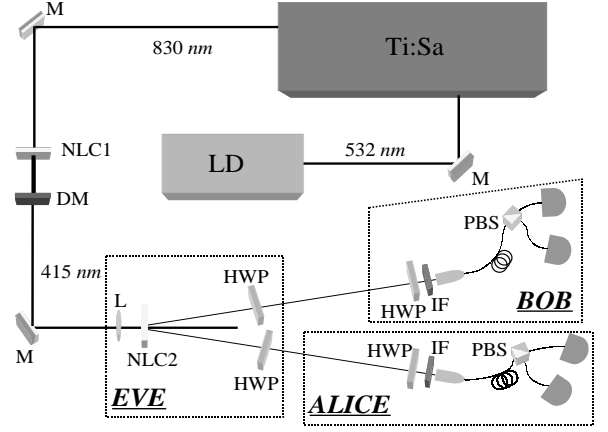}
\end{center}
\caption{ QKD set-up with the source of photon pairs under Eve's
control: photon pairs are generated by SPDC in a type I nonlinear
crystal (NLC2) pumped by the pulsed laser system (LD, Ti:Sa and
NLC1). The polarization state of the photons is controlled by
half-wave plates (HWPs) and selected photons are directed to the
Alice and Bob detection apparatuses composed of HWPs, interference
filters (IF) fiber couplers, fibers integrated polarizing beam
splitters (PBS), single-photon detectors. M mirror, DM dichroic
mirror, L lens} \label{Figure 3}
\end{figure}

In Fig. 2 we depict the experimental scheme considering the
situation in which Eve has total control of the source. In this
scheme, the source under Eve's control replaces the source of
entangled photon pairs of a typical QKD scheme
\cite{ekertrarity,sasha,qk2,qk3,qk1}. Eve's source is obtained by
a 1 mm length LiIO$_{3}$ nonlinear crystal (NLC2) pumped by
ultrashort pulses (150 fs) at 415 nm generated from a second
harmonic (obtained from NLC1) of a ultrashort mode-locked
Ti-Sapphire with a repetition rate of 76 MHz pumped by a 532 nm
green laser. The NLC2 realizes a non-collinear type I phase
matching and Eve selects two quantum correlated optical channels
along which the twin photons at 830 nm (emitted at 3.4$^{\circ}$)
are sent towards Alice and Bob's detection apparatuses
\cite{prl,klyshko}. The down-converted photons of a pair have the
same polarization state (ordinary waves) and Eve can modify
deterministically the polarization state of the photon by means of
a half-wave-plate (HWP) in each channel, in other words Eve sends
photon pairs to Alice and Bob with polarization state $\left|
\phi_{A}\right\rangle$ and $\left| \phi_{B}\right\rangle$,
respectively.

Alice and Bob's detection apparatuses are identical and are
composed of an open air-fiber coupler to collect the
down-converted light by single-mode optical fibers. The detection
of photons in the proper polarization basis is guaranteed by a HWP
before the fiber coupler and a fiber-integrated polarizing beam
splitter (PBS). Photons at the two output ports of the PBS are
sent to fiber coupled photon counters (Perkin-Elmer SPCM-AQR-14)
\cite{disc}. Interference filters peaked at 830 nm with 11 nm
bandwidth are placed in front of the fiber couplers to reduce
straylight.

Coincident counts between any of Alice's detectors
($+_{A},\,-_{A}$) and any of Bob's detectors ($+_{B},\,-_{B}$) are
obtained from an Elsag prototype of four-channel coincident
circuit \cite{elsag1,elsag2}. Single-counts and coincidences are
counted by a National Instruments \cite{disc} sixteen channels
counter plug-in PC card.

The terms $p_{\alpha _{A},\alpha _{B}}(x_{A},y_{B})$ are estimated
in terms of the number of coincident counts:
\begin{equation}
p_{\alpha _{A},\alpha _{B}}(x_{A},y_{B})=\frac{N_{\alpha
_{A},\alpha
_{B}}(x_{A},y_{B})}{%
\begin{array}{c}
\lbrack N_{\alpha _{A},\alpha _{B}}(+_{A},+_{B})+N_{\alpha
_{A},\alpha
_{B}}(+_{A},-_{B})+ \\
N_{\alpha _{A},\alpha _{B}}(-_{A},+_{B})+N_{\alpha _{A},\alpha
_{B}}(-_{A},-_{B})]
\end{array}
}
\end{equation}
where $N_{\alpha _{A},\alpha _{B}}(x_{A},y_{B})$ is the number of
coincidences measured by the couple of detectors $x_{A},\,y_{B}$
($x,y=+,-$) when Alice and Bob's detection apparatuses project
photons in the polarization bases at Eq.s (\ref{polbas}).

\begin{figure}[tbp]
%[htbp]
\par
\begin{center}
\includegraphics[angle=0, width=9 cm, height=5.5 cm]{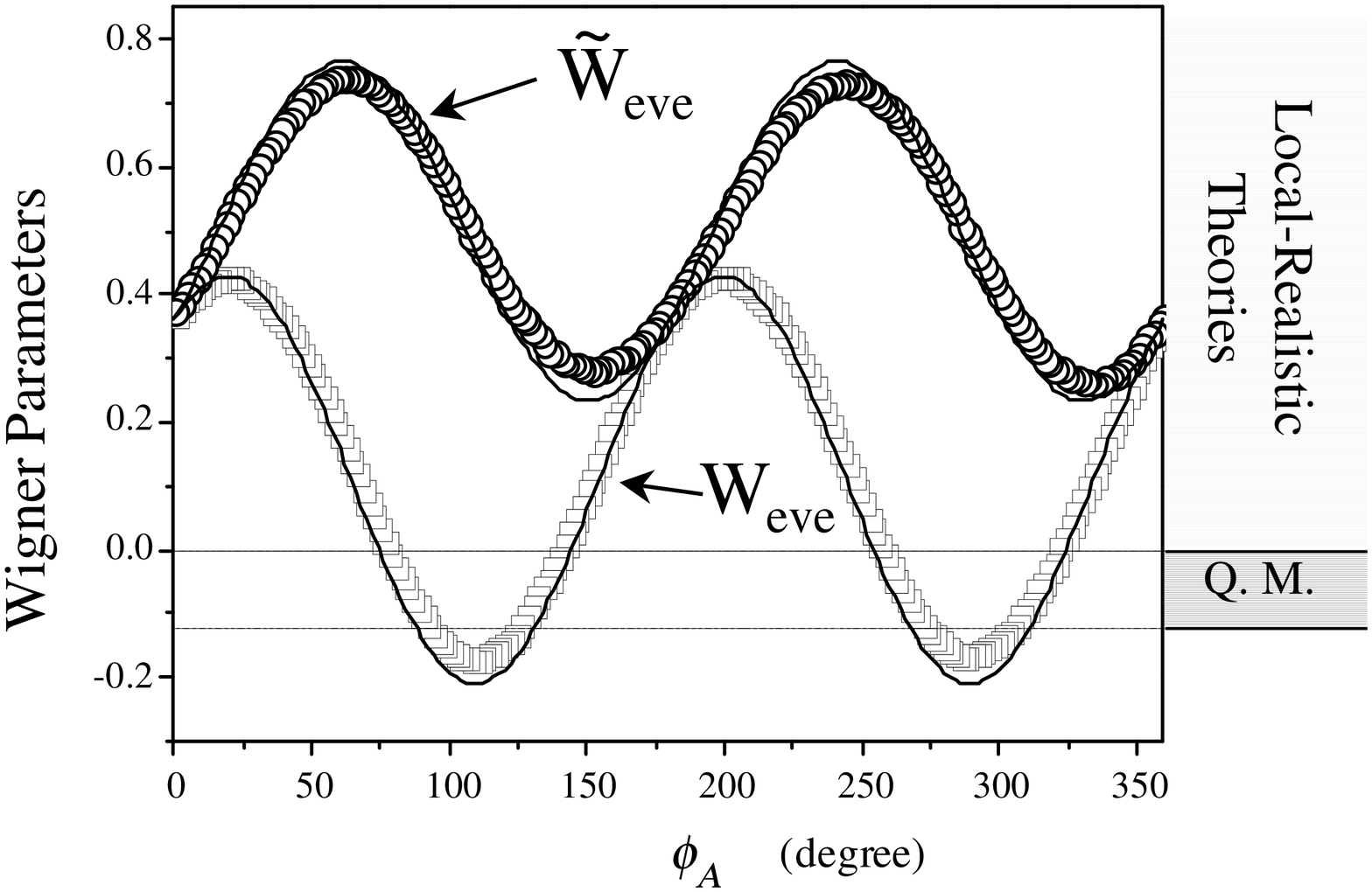} \\
\includegraphics[angle=0, width=9 cm, height=5.5 cm]{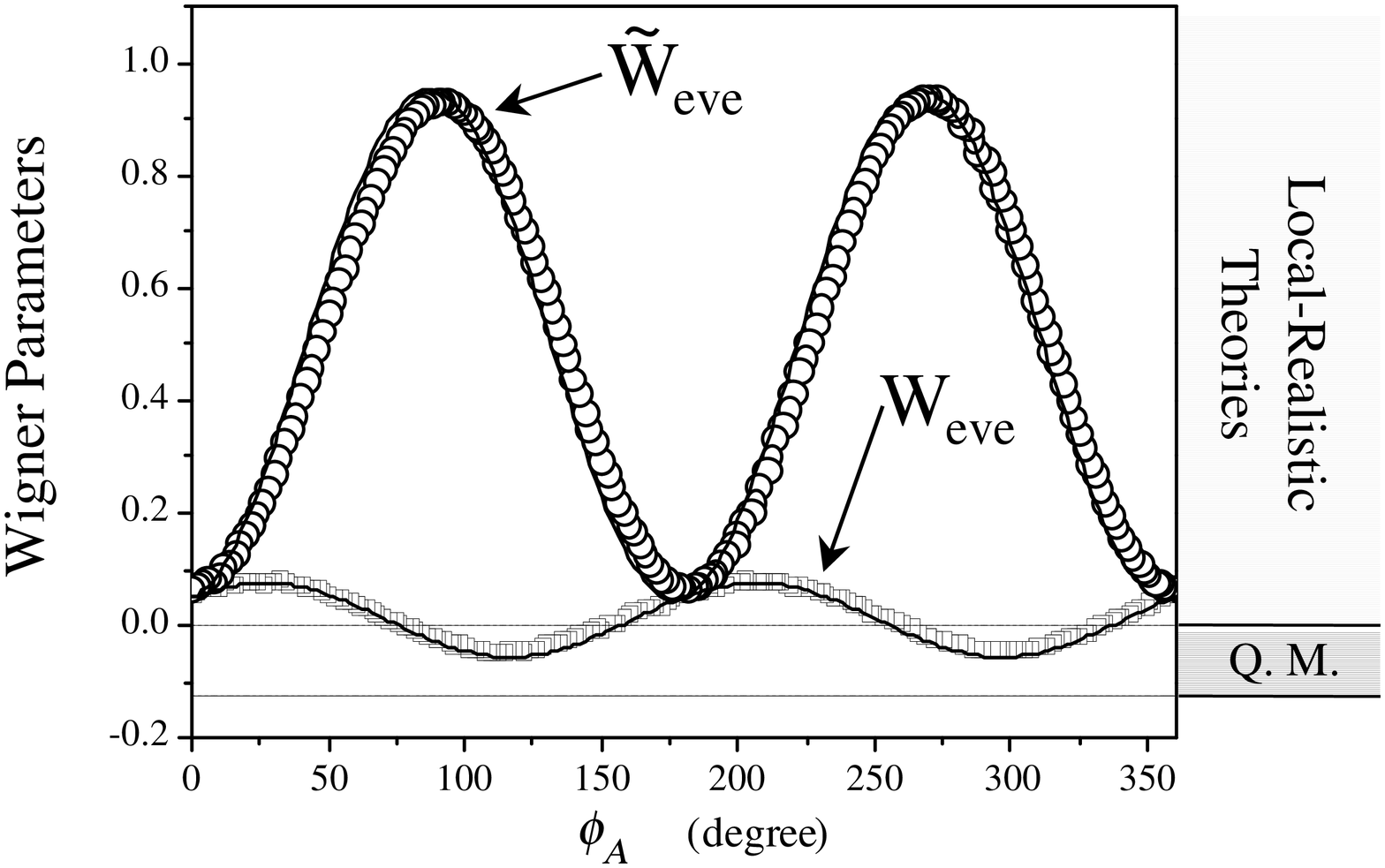} \\
\includegraphics[angle=0, width=9 cm, height=5.5 cm]{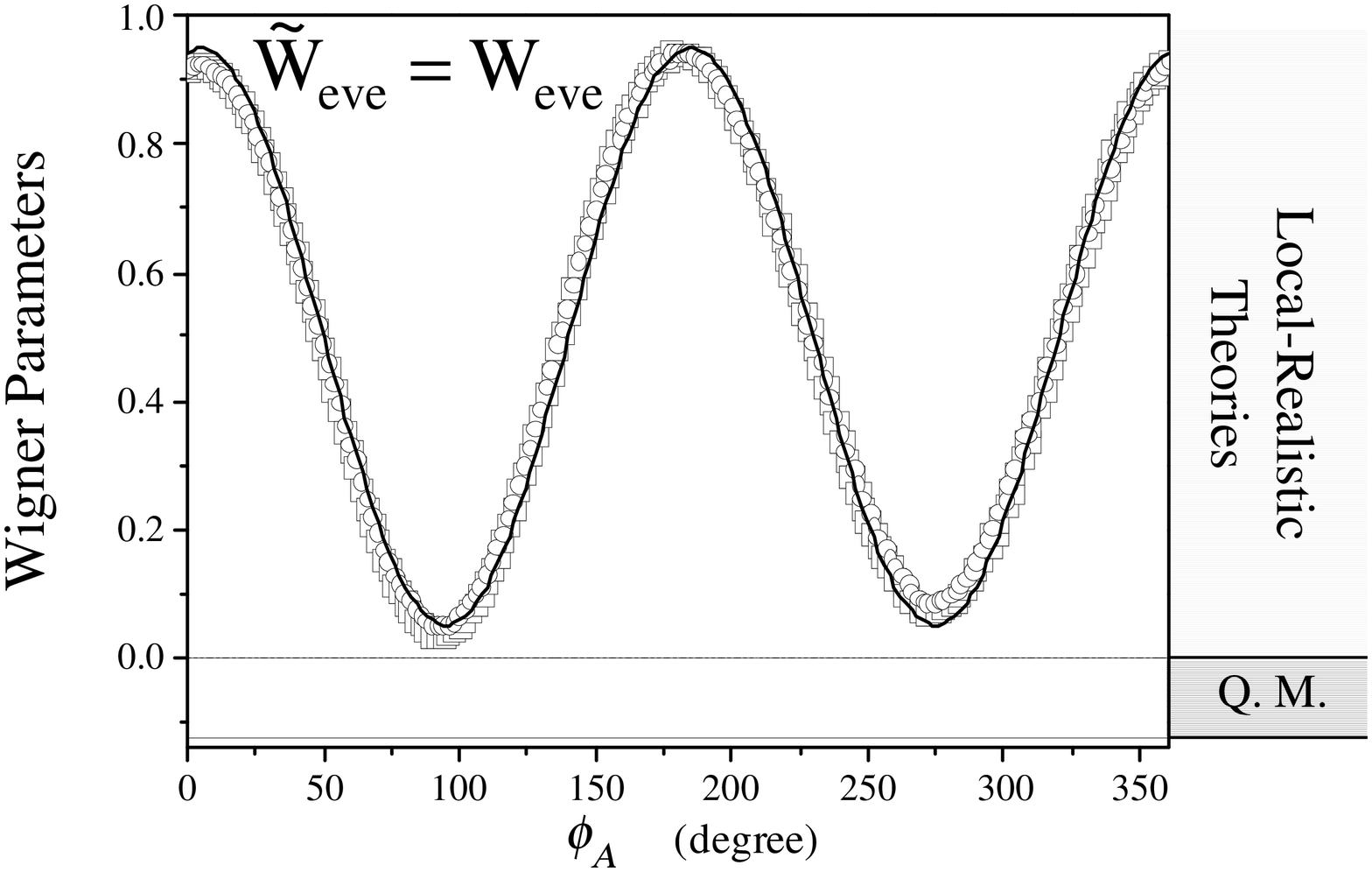}
\end{center}
\caption{ Experimental data for $W$ (squares) and $\widetilde{W}$
(circles) and theoretical curves (lines) are showed. Range of
values of Wigner's parameter corresponding to local-realistic
theories and quantum mechanics (Q.M.) are indicated. (a) Violation
of the limit  of local-realistic theories for $W$, obtained with
$\phi _{B}=62^{\circ}$.  (b) Minimum obtainable for
$\widetilde{W}$ obtained with $\phi _{B}=98^{\circ}$, along with
violation of the limit  of local-realistic theories for $W$. (c)
$\widetilde{W}=W$ obtained with $\phi _{B}=0^{\circ}$. }
\label{Figure 4}
\end{figure}

In Fig. 3 (a) we present our main result: photons sent by Eve in a
definite polarization state violate the limit  of local-realistic
theories. Experimental data for $W$ (circles) and $\widetilde{W}$
(squares) are presented versus $\phi _{A}$, with $\phi _{B}$ fixed
approximately at $62^{\circ}$ and show a good agreement with
theoretical predictions (lines).

As expected from the theory \cite{wignerien}, not only does $W$
violate the limit  of local-realistic theories ($W=0$), but also
some data points pass the quantum limit ($W=-0.125$); while
$\widetilde{W}$ is well above the limit of local-realistic
theories. The theoretical curves are calculated with $\phi
_{A}=62^{\circ}$, and the discrepancy between theory and
experiment can be explained by noting the difficulties in the
proper angular positioning of the four HWPs and in the noise
introduced by real optical devices, e.g., fibers, PBSs, detectors
dark counts and straylight.

In Fig. 3 (b) we present the experimental data and the theoretical
curve obtained with $\phi _{B}=98^{\circ}$, corresponding to a
position close to the minima of $\widetilde{W}$ as predicted by
the theory. Fig. 3 (b) shows a good agreement between experimental
data (circles) and theoretical predictions of $\widetilde{W}$ and
the minimum of experimental values, 0.0685, is slightly higher
than the theoretical predictions of 0.0466.

Furthermore, Fig. 3 (b) shows also the experimental data for $W$
(small squares) together with the associated theoretical curve,
and we observe that also in this case a violation of the
local-realistic theories limit occurs.

According to Eq.s (\ref{WIG1}) and (\ref{WIG2}), we note that
$\widetilde{W}$ differs from $W$ only because of the term
$p_{0^{\circ} _{A},0^{\circ} _{B}}(-_{A},-_{B})$, thus if
$p_{0^{\circ} _{A},0^{\circ} _{B}}(-_{A},-_{B})=0$  then
$\widetilde{W}=W$, and this occurs when $\phi _{A}=0^{\circ}$,
$180^{\circ}$ or $\phi _{B}=0^{\circ}$, $180^{\circ}$.

In Fig. 3 (c) we consider the situation when $\phi _{B}=0^{\circ}$
and we observe  that the experimental data for $\widetilde{W}$
(small circles) are almost superimposed to the $W$ ones (squares)
in good agreement with the theoretical prediction,
$\widetilde{W}=W$ (line).

Some further analysis of $\widetilde{W}$ must be considered for
the practical implementation of the Ekert protocol based on
Wigner's inequality. According to \cite{qk2}, we highlight that
the Ekert's protocol based on modified Wigner's inequality still
guarantees a simplification with respect to the one based on the
CHSH inequality, because Alice and Bob randomly choose between two
rather than three bases. Though the necessity of an experimental
evaluation of the term
$p_{0^{\circ}_{A},0^{\circ}_{B}}(-_{A},-_{B})$ forces Alice and
Bob to sacrifice part of the key for the sake of security, we note
that in any practical implementation of QKD protocols, Alice and
Bob distill from the noisy sifted key a nearly noise-free
corrected key by means of error correction procedures subjected to
the constraint of knowing the quantum bit error rate (QBER). Also,
the QBER is estimated at the cost of losing part of the key. Thus,
we suggest using the same sacrificed part of the key to estimate
both $\widetilde{W}$ and QBER.

To perform a proper comparison of the performances of Ekert
protocols based on Wigner's inequality versus the one CHSH-based
\cite{ekert}, it is necessary to consider situations where the
same number of analyzer settings are employed. In particular, we
consider the modified protocol based on Wigner inequality proposed
in \cite{wignerien} where Alice and Bob measure randomly using
three analyzer settings (as in the case of CHSH). This protocol is
more efficient than the protocol based on CHSH. In particular, for
CHSH only 2/9 of the qubits exchanged are devoted to the key
generation \cite{ekert}, while here we can improve till 1/3
depending on the security needs. Furthermore in this protocol none
of the qubits exchanged are discarded while in the case of CHSH
1/3 of the qubits are discarded \cite{wignerien}.

In conclusion, this paper highlights the insecurity of Ekert's
protocol based on the Wigner inequality. We performed an
experiment simulating the  total control of photons in Alice and
Bob channels by an eavesdropper, proving that the QKD Ekert
protocol based on Wigner's inequality presents a serious lack of
security. In addition, we proved that a modified version of the
Wigner security parameter guarantees secure QKD.

We are indebted to P. Varisco, A. Martinoli, P. De Nicolo, S.
Bruzzo, I. Ruo Berchera, G. Di Giuseppe. This experiment was
carried out in the Quantum Optics Laboratory of Elsag S.p.A.,
Genova (Italy), within a project entitled "Quantum Cryptographic
Key Distribution" co-funded by the Italian Ministry of Education,
University and Research (MIUR) - grant n. 67679/ L. 488. In
addition S. C. acknowledges the partial support of the DARPA QuIST
program and M. L. R. acknowledges the partial support by INFM.


\begin{thebibliography}{99}
\bibitem{disc}  Certain trade names and company products are mentioned in the text
or identified in an illustration in order to adequately specify
the experimental procedure and equipment used. In no case does
such identification imply recommendation or endorsement by the
Istituto Elettrotecnico Nazionale G. Ferraris, by ELSAG S.p.A., or
by National Institute of Standards and Technology nor does it
imply that the products are necessarily the best available for the
purpose.
\end{thebibliography}
\end{document}